# Cathodoluminescence spectroscopy of monolayer hexagonal boron nitride

K. Shima,[1,*] T. S. Cheng,[2] C. J. Mellor,[2] P. H. Beton,[2] C. Elias,[3] P. Valvin,[3] B. Gil,[3] G. Cassabois,[3] S. V. Novikov,[2] and S. F. Chichibu[1,*]

[1]Institute of Multidisciplinary Research for Advanced Materials, Tohoku University, Sendai 980-8577, Japan
[2]School of Physics and Astronomy, University of Nottingham, Nottingham NG7 2RD, UK
[3]Laboratoire Charles Coulomb, UMR5221 CNRS-Université de Montpellier, 34095 Montpellier, France

E-mail: chichibulab@yahoo.co.jp

**Abstract**

Cathodoluminescence (CL) spectroscopy is a powerful technique for studying emission properties of optoelectronic materials because CL is free from excitable bandgap limits and from ambiguous signals due to simple light scattering and resonant Raman scattering potentially involved in the photoluminescence (PL) spectra. However, direct CL measurements of atomically thin two-dimensional materials, such as transition metal dichalcogenides and hexagonal boron nitride (hBN), have been difficult due to the small excitation volume that interacts with high-energy electron beams (*e*-beams). Herein, distinct CL signals from a monolayer hBN, namely mBN, epitaxial film grown on a highly oriented pyrolytic graphite substrate are shown by using a home-made CL system capable of large-area and surface-sensitive excitation by an *e*-beam. The spatially resolved CL spectra at 13 K exhibited a predominant 5.5-eV emission band, which has been ascribed to originate from multilayered aggregates of hBN, markedly at thicker areas formed on the step edges of the substrate. Conversely, a faint peak at 6.04±0.01 eV was routinely observed from atomically flat areas. Since the energy agreed with the PL peak of 6.05 eV at 10 K that has been assigned as being due to the recombination of phonon-assisted direct excitons of mBN by Elias *et al*. [Nat. Commun. **10**, 2639 (2019)], the CL peak at 6.04±0.01 eV is attributed to originate from the mBN epilayer. The CL results support the transition from indirect bandgap in bulk hBN to direct bandgap in mBN, in analogy with molybdenum disulfide. The results also encourage to elucidate emission properties of other low-dimensional materials with reduced excitation volumes by using the present CL configuration.

Two-dimensional (2D) layered materials, such as graphite, hexagonal boron nitride (hBN), and transition metal dichalcogenides (TMDs), are the building blocks of van der Waals (vdW) heterostructures[1,2] that are promising platform for optoelectronics[3] and valleytronics.[4,5] An isolation of monolayer 2D materials causes unique optoelectrical phenomena: Mak *et al*.[3] have demonstrated a transition from indirect bandgap in bulk molybdenum disulfide ($MoS_2$) to direct bandgap in monolayer $MoS_2$, where the monolayer $MoS_2$ exhibited an increased luminescence quantum efficiency by more than a factor of $10^4$ compared with the bulk $MoS_2$.[3] For realizing vdW heterostructures consisting of different 2D materials with desired band alignment and interlayer coupling, it is essential to understand the fundamental optoelectronic properties of 2D materials.

hBN crystallizes in layers of a 2D honeycomb structure based on in-plane three $sp^2$ covalent bonds that are connected by out-of-plane π bonds. Accordingly, hBN is a key building block in vdW heterostructures based on graphite,[2] because of the large bandgap energy ($E_g$) of 6 eV[6,7] and a small lattice mismatch (~1.8 %) to graphite. On another front, hBN is an exotic candidate for the use in deep ultraviolet (DUV) light emitters despite its indirect bandgap:[6] a lasing action at 5.77 eV has been reported by Watanabe *et al*. in 2004 from hBN single crystals grown by high-pressure high-temperature synthesis,[8] followed by the operation of planar DUV light-emitting device.[9] With respect to condensed matter physics of hBN, Cassabois *et al*.[6] have revealed that hBN has an indirect bandgap with the nonphonon (NP) indirect exciton (iX) energy of 5.955 eV at 10 K. In bulk hBN, iXs built from **M** and **K** points of the Brillouin zone (BZ) for the conduction and valence bands, respectively,[10-13] require the scattering by phonons[14,15] of wavevector **MK** to fulfill momentum conservation during photon absorption or emission. Nevertheless, the near-band-edge (NBE) emission of hBN, namely LO(T) and TO(T) phonon-assisted iXs [iX$_{LO(T)}$ and iX$_{TO(T)}$, respectively],[6] where T indicates **T** point of the BZ, exhibited markedly high internal quantum efficiency of 50 % at 10 K.[7] Such a high internal quantum efficiency has been explained from the macroscopic degeneracy of the parallel indirect transitions between the flat bands along the **KH** and **ML** lines of the Brillouin zone.[16]

In contrast to bulk hBN, very little is known about the optical transition properties of hBN of reduced layer numbers. Theoretical calculations have predicted a direct bandgap at **K** point for a monolayer hBN (mBN)[12,17] and an indirect bandgap or marginally direct bandgap for stackings of two or more layers,[17-19] in analogy with MoS$_2$.[3] Elias *et al*.[20] have confirmed the presence of the direct bandgap with $E_g$ of 6.1 eV in the mBN epilayers grown by high-temperature plasma-assisted molecular beam epitaxy (HT-PAMBE) method on a highly oriented pyrolytic graphite (HOPG) substrate,[21-23] by means of optical reflectance (OR) and photoluminescence (PL) measurements: the PL spectrum of mBN at 10 K has consisted of doublet peaks at around 6.08 eV and 6.05 eV, which were interpreted as the recombination of an NP direct exciton (dX), namely dX$_{NP}$, and a ZA(K) phonon[24,25]-assisted dX [dX$_{ZA(K)}$], respectively.[20] The 6.1-eV emission peaks associated with the recombination of dXs have been additionally verified by the optical probing techniques.[26-29]

For the accurate understanding of a luminescence spectrum of mBN, the complementary use of cathodoluminescence (CL) measurements is preferred because an electron beam (*e*-beam) excitation is free from excitable $E_g$ limits and from ambiguous signals due to simple light scattering and resonant Raman scattering potentially involved in PL spectra.[20] Schué *et al*.[30] have carried out conventional CL measurements on exfoliated hBN flakes to record a series of CL spectra as a function of the number of hBN layers from 100 down to 6. Their CL spectra[30] exhibited the NBE emissions at 10 K with the highest energy peak at 5.9 eV, which was followed by several phonon replicas. They[30] also observed a thickness-dependent energy shift of the 5.9 eV peak by a few tens of meV.[30] However, CL signals of the hBN flake thinner than 5 layers have not been shown yet due to the small excitation volume of ultrathin hBN films including mBN and to the finite depth of the projected range of generally focused *e*-beam in the CL system equipped on a scanning electron microscopy (SEM), which results in surface-insensitive excitation. We note that, for the cases of monolayer TMDs, the CL signals have been recorded only when the monolayer TMDs have been encapsulated with hBN layers, where the hBN has increased the gross excitation volumes and the excited carriers in hBN have been transferred to the monolayer TMDs.[31-34] However, such artificial vdW structures[31-34] are not feasible when an mBN is the active layer.

In this letter, distinct CL signals from the mBN epilayer[21-23] grown by HT-PAMBE on an HOPG substrate are shown. For overcoming the drawbacks of conventional CL measurement, a home-made CL system[35] that is capable of large-area and surface-sensitive excitation by an *e*-beam was used. The CL spectra at 13 K exhibited a predominant 5.5-eV emission band and a faint peak at 6.04±0.01 eV. Since the latter energy agreed with the PL peak of 6.05 eV (10 K) that has been assigned as being due to the recombination of dX$_{ZA(K)}$ in mBN,[20] the CL peak at 6.04±0.01 eV most likely originates from mBN. The result supports the direct bandgap of 6.1 eV in mBN.[20] With respect to the 5.5-eV band, which has been ascribed to originate from multilayered aggregates of hBN,[20,21,23] spatially resolved cathodoluminescence (SRCL) measurement revealed that the emissions were localized markedly at thicker areas formed on step edges of the HOPG substrate.

A HT-PAMBE system[36] was used to grow the mBN epilayer on a 10×10-mm$^2$-area HOPG substrate with a mosaic spread of 0.4°.[21-23] To obtain a fresh graphite surface for the vdW epitaxy, the top surface of the HOPG substrate was removed by exfoliation using an adhesive tape. After the exfoliation, the HOPG substrate was cleaned with toluene to remove any remaining tape residue, followed by an annealing at 200 ºC for 4 hours in a mixed gas ambient of Ar and H$_2$ (5 %). To supply a boron flux, a high-temperature effusion cell (Veeco) containing a high-purity (99.999%) natural mixture of $^{11}$B and $^{10}$B isotopes was heated up to 1875 ºC. To supply an active nitrogen flux, an RF plasma source (Veeco) was used with the power of 550 W and an N$_2$ flow rate of 2 sccm. The mBN epilayer was grown at 1390 ºC for 3 h. The growth parameters were identical to the mBN epilayers that have been characterized by using OR and PL measurements.[20] Atomic force microscopy measurements[20,28] revealed that overall surface coverage of the grown BN film was approximately 100 %, with approximately 89 % of the surface covered predominantly by mBN, together with some small regions of multilayered hBN deposits that nucleated at step edges of the HOPG substrate. No chemical intermixing was observed between the mBN epilayer and the HOPG substrate.[21,37]

Wide-area CL measurement was carried out on the mBN epilayer at low temperatures using a homemade CL system,[35]

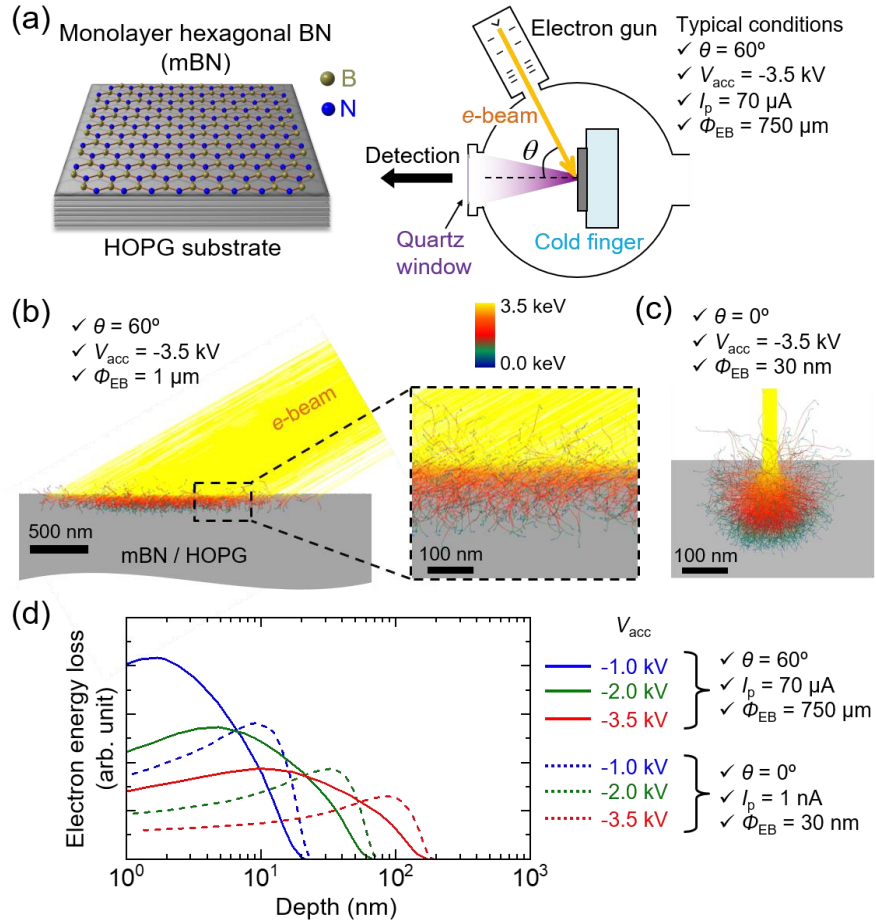

**Figure 1.** (a) Schematic representation of the wide-area cathodoluminescence (CL) measurement of monolayer hexagonal BN (mBN) epilayer grown on a highly oriented pyrolytic graphite (HOPG) substrate. The typical incident angle ($\theta$), acceleration voltage ($V_{acc}$), probe current ($I_p$), and diameter ($\phi_{EB}$) of the *e*-beam were 60º, -3.5 kV, 70 µA, and 750 µm, respectively. Monte Carlo simulations of the electron trajectories into a model mBN / HOPG for (b) the wide-area CL and (c) the spatially resolved CL (SRCL) measurements. They were calculated using the CASINO software.[32] In the simulation, $\theta$, $V_{acc}$, and $\phi_{EB}$ were set respectively to 60º, -3.5 kV, and 1 µm for the wide-area CL and 0º, -3.5 kV, and 30 nm for the SRCL measurements. (d) Simulated depth distributions of the total energy loss of irradiated electrons into mBN / HOPG for the wide-area CL (solid lines) and the SRCL (dashed lines) measurements with $V_{acc}$ = -1.0, -2.0, and -3.5 kV.

as schematically shown in the right panel of Fig 1(a). In order to increase the gross excitation volume of an ultrathin film using an *e*-beam, large-area and surface-sensitive excitation was realized by adjusting the incident angle ($\theta$), acceleration voltage ($V_{acc}$), and probe current ($I_p$) of the *e*-beam to 60º, -3.5 kV, and 70 µA, respectively. As a result, an *e*-beam diameter ($\phi_{EB}$) of approximately 750 µm and probe current density ($J_p$) of 16 mA·cm$^{-2}$ were obtained. These parameters were varied to study the influences on the CL intensities from the mBN epilayer. Conventional SRCL measurements were carried out at low temperatures using the system equipped on the SEM (JSM-6510) under the conditions of $\theta$ = 0º, $V_{acc}$ = -2.0 kV, and $I_p$ = 8 nA, giving $\phi_{EB}$ of approximately 100 nm. The dimensions of excitation volumes for the mBN / HOPG structure were calculated using a Monte Carlo simulator, CASINO software.[38] Figures 1(b) and 1(c) show the simulated electron trajectories into a model mBN / HOPG structure for the wide-area CL and SRCL measurements, respectively. In the simulation, $\theta$, $V_{acc}$, and $\phi_{EB}$ were set respectively to 60º, -3.5 kV, and 1 µm for the wide-area CL and 0º, -3.5 kV, and 30 nm for the SRCL measurements. The simulated depth distributions of the total energy loss of irradiated electrons into mBN / HOPG structure with $V_{acc}$ = -1.0, -2.0, and -3.5 kV for the wide-area CL and SRCL measurements are shown by solid and dashed lines, respectively, in Fig. 1(d). Compared with the SRCL measurement, our wide-area excitation appears to enable more surface-sensitive CL measurement, which contributes to increase gross CL intensity of the ultrathin mBN epilayer.

A wide-area CL spectrum at 13 K of the mBN epilayer is shown by blue solid line in Fig. 2(a). For comparison, wide-area CL spectra of a bare HOPG substrate (13 K, gray solid line) and approximately 1-μm-thick hBN epilayer (12 K, green solid line) that was grown on a (0001) sapphire by low-pressure chemical vapor deposition (LP-CVD) using a BCl$_3$-NH$_3$-N$_2$ gas system[39-41] are also displayed. The CL spectrum of the mBN consisted of a predominant multi-peaked broad emission band at around 5.5 eV and a faint shoulder due to certain independent emission peak at approximately 6.04 eV. Only stray light was recorded in the reference spectrum of the bare HOPG, giving a proof that the 5.5-eV band and 6.04-eV peak originate from the mBN epilayer. The 5.5-eV band[42-47] has been found in hBN samples[41-48] and assigned as the emissions of iX$_{LO(T)/TO(T)}$ further scattered by multiple TO(K) phonons [iX$_{LO(T)/TO(T)+n\text{TO(K)}}$, $n$ is an integer][47] and other iXs trapped by certain stacking defects.[44-47] The 5.5-eV band has also been found in the PL spectrum at 10 K of the same series[20] of mBN / HOPG grown by HT-PAMBE. The origin of the 5.5-eV band will be discussed later. The energy of the 6.04-eV CL peak nearly agreed with the PL peak at 6.05 eV (10 K) of the same series[20] of mBN samples, where the PL peak has been assigned as being due to the recombination of dX$_{ZA(K)}$ in mBN in accordance with the results of OR and PL measurements.[20] Meanwhile, the CL spectrum of the LP-CVD hBN film exhibited three emission groups labeled "NBE emissions", "5.5-eV band", and "4.0-eV band".[41] Among these, the origin of the 4.0-eV band[40,42,48-54] has been suggested to associate with carbon and oxygen impurities and a nitrogen vacancy (V$_N$).[42,48,50-53] We note that the NBE emissions of the LP-CVD hBN was accompanied by a distinct emission peak at around 6.035 eV at 12 K that has been detected from polytypic segments,[41] most probably graphitic bernal BN (bBN).[55-57] As shown in Fig. 2(a), the CL intensity for the NBE emission of mBN at 6.04 eV was less than three (less than four) orders of magnitude lower than that of the ZA(T) phonon-assisted iXs [iX$_{ZA(T)}$][6] at 5.92 eV (iX$_{TO(T)}$ at 5.79 eV) of the LP-CVD hBN film.[39-41] Because the thickness of mBN (i.e., 0.3-0.4 nm)[21] was approximately 3000 times smaller than that of the LP-CVD hBN film (i.e., 1 μm),[39-41] external quantum efficiencies for the present mBN and the LP-CVD hBN film[39-41] appear to be the same order of magnitude.

The NBE CL spectra at 13 K of the mBN epilayer are shown in the bottom panel of Fig. 2(b). For reference, the OR and PL spectra of the same series of mBN / HOPG measured at 10 K by Elias *et al*.[20] are reproduced in the top panel of Fig.

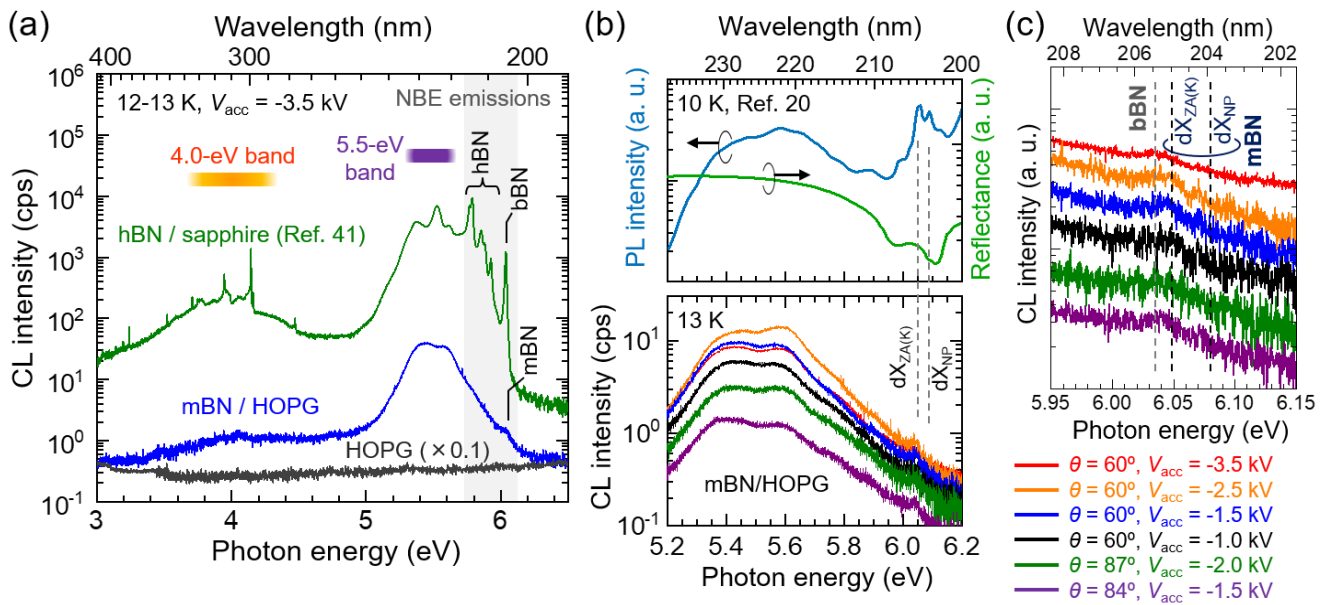

**Figure 2.** A wide-area CL spectrum at 13 K of the mBN epilayer (blue solid line). For comparison, wide-area CL spectra of a bare HOPG substrate (13 K, gray solid line) and 1-μm-thick hBN epilayers[41] (12 K, green solid line) that was grown on a (0001) sapphire by low-pressure chemical vapor deposition using a BCl$_3$-NH$_3$-N$_2$ gas system are also displayed. $\theta$, $V_{acc}$, $I_p$, and $\phi_{EB}$ of the *e*-beams were 60º, -3.5 kV, 70 μA, and 750 μm, respectively. NBE stands for near-band-edge. [Partially reproduced with permission from Ref. 41, Appl. Phys. Lett. **120**, 231904 (2022). Copyright 2022 AIP Publishing LLC]. (b) The NBE CL spectra at 13 K of the mBN epilayer measured under various $\theta$ and $V_{acc}$. For reference, OR and PL spectra[20] of the same series[20] of mBN / HOPG measured at 10 K are displayed. [Partially reproduced with permission from Ref. 20, Nat. Commun. **10**, 2639 (2019). Copyright 2010 Springer Nature Limited]. dX$_{NP}$ and dX$_{ZA(K)}$ stand for a nonphonon (NP) direct exciton (dX) and a ZA(K) phonon-assisted dX, respectively. (c) An enlarged view of the CL spectra between 5.95 and 6.15 eV of panel (b). The CL spectra are vertically offset for better visibility.

2(b). The values of $\theta$ and $V_{acc}$ for the CL measurement were varied to maximize overall emission intensity from the mBN epilayer. Regardless of excitation conditions, the mBN epilayer exhibited a distinct CL peak at around 6.04 eV. As described in the preceding paragraph, the peak energy nearly agreed with that of $dX_{ZA(K)}$ of mBN (6.05 eV at 10 K).[20] However, the other PL peak of 6.08 eV at 10 K that has been attributed to $dX_{NP}$ of mBN[20] was not clearly observed in the present CL spectra, potentially due to insufficient intensity. The intensity of the CL peak at 6.04 eV exhibited a maximum under the condition of $\theta = 60°$ and $V_{acc} = -3.5$ kV, of which spectrum is drawn by an orange solid line in Fig. 2(b). Further higher $\theta$ up to 87° resulted in lower CL intensities presumably due to the reflection of the irradiated *e*-beam at or right below the surface. Further lower $V_{acc}$ down to -1.0 kV also resulted in lower CL intensities, because the *e*-beam was less converged and gave lower $J_p$.

The CL spectra between 5.95 and 6.15 eV of Fig. 2(b) are enlarged in Fig. 2(c), where the spectra are vertically offset for better visibility. The NBE emission of the mBN epilayer was observed at 6.04±0.01 eV in all the spectra. This energy range again coincide with the $dX_{ZA(K)}$ of mBN.[20] Although the energy of the CL peak at 6.04±0.01 eV is close to that of the bBN emission at 6.035 eV at 12 K,[41] such possibility is ruled out because bBN is built from at least two mBN layers with AB stacking[56] while approximately 89 % of the surface of our sample was covered by mBN. Consequently, the CL peak at 6.04±0.01 eV at 13 K can be assigned as being due to the recombination of $dX_{ZA(K)}$.[20] Meanwhile, a CL peak for the recombination of $dX_{NP}$ at approximately 6.08 eV, which has been assigned by Elias *et al.* using PL measurements,[20] was not clearly observed, probably due to the insufficient CL intensity. The less pronounced emission intensity of the recombination of $dX_{NP}$ compared with that of $dX_{ZA(K)}$ is consistent with the results of PL measurements,[20] which likely arose from strong exciton-phonon interaction.[20,28] Nevertheless, the present CL results strongly support the direct bandgap of 6.1 eV for mBN, which has been determined in Ref. 20 based on the OR and PL measurements, thanks to the fact that CL is free from an excitable $E_g$ limits and from ambiguous signals due to simple light scattering and resonant Raman scattering occasionally observed in PL spectra.

The spatial distribution of the 5.5-eV band intensity in the mBN epilayer was studied by conventional SRCL measurement using low $V_{acc}$, as follows. One of the origins of 5.5-eV band is iXs trapped by certain stacking defects.[27,44-47] Such stacking defects can be created in multilayered aggregates of hBN formed on step edges of the HOPG substrate.[20,21,23] Figure 3(a) shows the surface topographic (upper) and corresponding phase (bottom) images[28] of the same series of mBN / HOPG epilayer.[20] Brighter regions in the upper image corresponded to topographically higher regions due to aggregation of BN sheets at step edges of HOPG, while white regions in the bottom image corresponded to partially exposed surface of the HOPG substrate.[20,21,23] The surface coverage of the multilayered aggregates (hBN) in the present epilayer was estimated to be approximately 11 %, where the aggregates were located at every a few micrometers. In order to correlate the locations of the aggregates and local CL spectra, spot-excitation CL measurements were carried out at the positions labeled P1 to P10 in the SEM image of the mBN epilayer, as shown in Fig. 3(b). The values of $\theta$, $V_{acc}$, $I_p$, and $\phi_{EB}$ were 0º, -2.0 kV, 8 nA, and approximately 100 nm, respectively. The local CL spectra are shown in Fig. 3(c), where the spectra are vertically offset for better visibility. As shown, the local CL spectra exhibited noisy line shapes with low S/N ratios, because the emission intensities of mBN (and partial hBN) under the SRCL measurement were far lower than those of wide-area CL measurement due to quite smaller excitation volume and less surface-sensitive excitation by the *e*-beam, as shown in Figs. 1(b)-1(d). Nevertheless, a distinguishable 5.5-eV band was found only in the spectrum for the position P6 (red line). In Fig. 3(d), the intensity profile of the 5.5-eV band along the positions labeled P1 to P10 in Fig. 3(b) is displayed. As shown, the 5.5-eV band emission was spatially distributed with an appearance rate of approximately 10 %, which is consistent with the surface coverage rate of approximately 11 % of the multilayered aggregates of hBN in the present mBN epilayer.

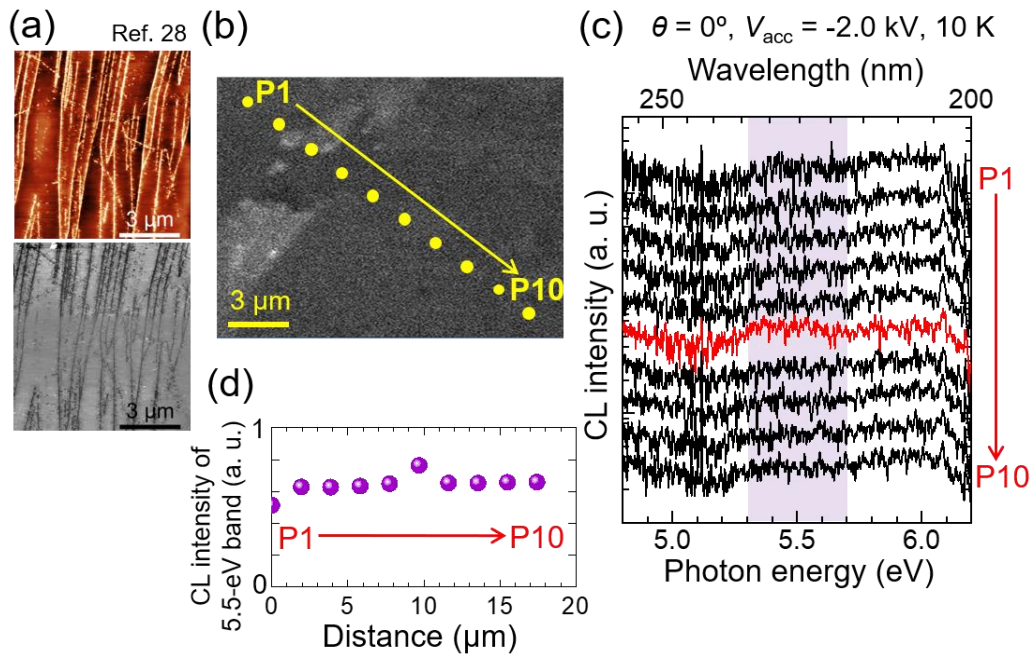

**Figure 3.** (a) Atomic force microscopy surface topographic (upper) and corresponding phase (bottom) images,[28] and (b) an SEM image of the mBN epilayer. [Partially reproduced with permission from Ref. 28, Phys. Rev. X **12**, 011057 (2022). Copyright 2022 American Physical Society]. (c) Local CL spectra at 10 K measured at the positions labeled P1-P10 in panel (b). $\theta$, $V_{acc}$, $I_p$, and $\phi_{EB}$ of the *e*-beams were 0°, -2.0 kV, 8 nA, and approximately 100 nm, respectively. The CL spectra are vertically offset for better visibility. (d) CL intensity profile of the 5.5-eV band along the positions labeled P1-P10 in panel (b). The spectral integration was carried out at the photon energies ($h\nu$) between 5.3 and 5.7 eV.

In summary, distinct CL signals were recorded from an mBN epilayer grown on an HOPG substrate by using a home-made CL system capable of large-area and surface-sensitive excitation by an *e*-beam. The CL spectra at 13 K exhibited a predominant 5.5-eV emission band, which has been ascribed to originate from multilayered aggregates of hBN, markedly at thicker areas formed on step edges of the substrate. Conversely, a faint peak at 6.04±0.01 eV was routinely observed from atomically flat areas. Since the energy agreed with the PL peak of 6.05 eV at 10 K that has been assigned as being due to the recombination of phonon-assisted direct excitons of mBN,[20] the CL peak at 6.04±0.01 eV most likely originates from the mBN epilayer. The results encourage to elucidate emission properties of mBN and other low-dimensional materials by using the present CL system.

## Acknowledgements

This work was supported in part by "Crossover Alliance to Create the Future with People, Intelligence, and Materials" and JSPS KAKENHI (Grant Nos. JP16H06427, JP17H02907, JP20K20993, and JP22H01516) by Ministry of Education, Culture, Sports, Science and Technology (MEXT), Japan. This work was also supported in part by the Engineering and Physical Sciences Research Council UK (Grant Nos. EP/K040243/1, EP/P019080/1, and EP/V05323X/1).